\documentclass{PoS}
\usepackage{amssymb}
\usepackage{graphicx}

\def\simleq{\; \raise0.3ex\hbox{$<$\kern-0.75em \raise-1.1ex\hbox{$\sim$}}\; }
\def\simgeq{\; \raise0.3ex\hbox{$>$\kern-0.75em \raise-1.1ex\hbox{$\sim$}}\; }

\newcommand{\GeV}{{\rm GeV}}
\newcommand{\TeV}{{\rm TeV}}
\newcommand{\PeV}{{\rm PeV}}

\newcommand{\cm}{{\rm cm}}

\newcommand{\s}{{\rm s}}
\newcommand{\sr}{{\rm sr}}

\title{A Hadronic Scenario for the Galactic Ridge}

\ShortTitle{Short Title for header}

\ShortTitle{Gamma-ray and Neutrino Galactic emissions above the TeV}

\author{Daniele Gaggero\\
        {SISSA and INFN, via Bonomea 265, I-34136 Trieste, Italy}\\
        E-mail: \email{daniele.gaggero@sissa.it}}
        
\author{Dario Grasso\\
        Dipartimento di Fisica ``E. Fermi" , Pisa University and I.N.F.N.,  Largo B. Pontecorvo 3, I-56127 Pisa, Italy\\
        E-mail: \email{dario.grasso@pi.infn.it}}
        
\author{\speaker{Antonio Marinelli}\thanks{A footnote may follow.}\\
        Dipartimento di Fisica ``E. Fermi", Pisa University and I.N.F.N., Largo B. Pontecorvo 3, I-56127 Pisa, Italy\\
        E-mail: \email{antonio.marinelli@pi.infn.it}}        

\author{Alfredo Urbano\\
        {SISSA and INFN, via Bonomea 265, I-34136 Trieste, Italy}\\
        E-mail: \email{alfredo.urbano@sissa.it}} 

\author{Mauro Valli\\
        {SISSA and INFN, via Bonomea 265, I-34136 Trieste, Italy}\\
        E-mail: \email{mauro.valli@sissa.it}}

\abstract{
Several observations from Fermi-LAT, up to few hundred GeV, and from H.E.S.S., up to $\sim 10$ TeV, reported an intense $\gamma$-ray emission from the inner part of the Galactic plane. 
After the subtraction of point-like contributions, the remaining $\gamma$-ray spectrum can provide important hints about the cosmic-ray (CR) population in that region.
In particular, the diffuse spectrum measured by both Fermi-LAT and H.E.S.S. in the Galactic Ridge is significantly harder with respect to the rest of the Galaxy. These results were recently interpreted in terms of a comprehensive CR transport model which, adopting a spatial dependent diffusion coefficient and convective velocity, reproduces Fermi-LAT results on the whole sky as well as local CR spectra.  
We showed as that model predicts a significantly harder neutrino diffuse emission compared to conventional scenarios: The predicted signal is able to account for a significant fraction of the astrophysical flux measured by IceCube.
In this contribution, we use the same setup to calculate the expected neutrino flux from several windows in the inner Galactic plane and compare the results with IceCube observations and the sensitivities of Mediterranean neutrino telescopes. 
In particular, for the ANTARES experiment, we compare the model expectations with the upper limits obtained from a recent unblinded data-analysis focused on the galactic ridge region. 
Moreover, we also show the expectations from the galactic ridge for the future KM3NeT observatory, whose position is optimal to observe this portion of the sky.
}

\FullConference{The 34th International Cosmic Ray Conference,\\
		30 July- 6 August, 2015\\
		The Hague, The Netherlands}

\begin{document}

\section{Introduction}

In 2006, the High Energy Stereoscopic System (H.E.S.S.) reported the observation of a $\gamma$-ray emission \cite{2006Natur.439..695A} from the Galactic Ridge (GR) region: $-0.8^{\circ}<l<0.8^{\circ}$, $|b|<0.3^{\circ}$ in Galactic coordinates. 
After the subtraction of the point-like component associated to Sagittarius A \cite{2004A&A...425L..13A} and the SNR G0.9+0.1\cite{2005A&A...432L..25A}, the spectrum of the diffuse emission was fitted using a power-law with index $\Gamma=-2.29\pm0.07_{stat}\pm0.20_{sys}$. 
The hard slope of the emission and its positional coincidence with the Central Molecular Zone (CMZ) cloud complex in the central 200 pc of the Milky Way led the H.E.S.S. collaboration to argue in favour of its hadronic origin. 
Nevertheless, the spectrum measured by H.E.S.S. is significantly harder than expected from $\pi_0$-decay if the CR spectral shape in the GC region is assumed to be similar to the locally observed one, as assumed in all conventional CR transport models. 
%More recently the Fermi-LAT collaboration modeled the GeV excess in the inner part of the galaxy with a superposition of a spherically-symmetric template and a ridge-like template\cite{2014PhRvD..89f3515M} consistent with the position of the CMZ. 
%
In principle such discrepancy may be just a local effect due to the proximity between the dense molecular gas with some CR sources located in the complex GC region \cite{2006Natur.439..695A}, although no compelling evidences have been provided so far in favour of that interpretation.  
On the other hand, a leptonic origin of the GR emission cannot be excluded a priori, as it may arise from bremsstrahlung of ultra relativistic electrons onto the molecular gas \cite{2013ApJ...762...33Y,2014PhRvD..89f3515M}.
 
We recently showed that the Fermi-LAT data suggest a different interpretation.  
Our scenario is based on a new phenomenological model proposed in Ref. \cite{2015PhRvD..91h3012G} and implemented with the {\tt DRAGON} code \cite{Dragonweb}. That model assumes a radial dependence for both the rigidity scaling index $\delta$ of the CR diffusion coefficient and the convective wind. 
This setup was introduced in order to solve the discrepancy between conventional models predictions and the spectrum of the $\gamma$-ray diffuse emission measured by Fermi in the inner GP region \cite{FermiLAT:2012aa}.
Since the model assumes that $\delta$ {increases} with the galactocentric radius $R$, it predicts a hardening of the CR propagated spectrum hence of the $\gamma$-ray diffuse emission in the inner Galaxy. 
   
In Ref. \cite{2015arXiv150400227G} we showed that, if the CR hardening measured by PAMELA and AMS-02 at about $250$ GV is taken into account in this setup (hereafter, KRA$_{\gamma}$ model), it is possible to consistently reproduce both the $\gamma$-ray measurements by Fermi-LAT and those taken by several ground experiments above the TeV. 
This is the case of the $\gamma$-ray emission measured by Milagro experiment in the region ($30^{\circ}< l <65^{\circ}$,$|b|<2^{\circ}$) with a median energy of 15 TeV, solving a long standing problem faced by conventional models \cite{MilagroPlane}.
In this work we show as the KRA$_{\gamma}$ model also consistently reproduces Fermi-LAT and H.E.S.S. data in the GR region (see also \cite{Dario_talk}). In both regions the origin of the emission is dominated by hadronic scattering. 

The fact that the KRA$_{\gamma}$ scenario predicts a significantly harder CR spectrum in the inner Galaxy has also relevant consequences for neutrino astronomy and may change the standard predictions regarding the diffuse neutrino emission (see {\it e.g.} \cite{Evoli:2007iy}). %since it predicts a diffuse emission of the Galaxy significantly larger and harder than that computed on the basis of conventional models (see {\it e.g.} \cite{Evoli:2007iy}).
This is very interesting in light of the recent IceCube detection of $37$ neutrino events with energy above $30~\TeV$ corresponding to a $5.7~\sigma$ excess on top of the atmospheric background \cite{Aartsen:2014gkd}. 
The inferred flavour composition is compatible with a mixture of electronic, muonic and tauonic neutrino in equal amounts as expected if their origin were astrophysical. 
Recent analyses \cite{Aartsen:2014muf, 2015arXiv150703991I} based on neutrino events with vertices contained in the detector allowed the IceCube collaboration to lower the energy threshold and to measure the extraterrestrial diffuse neutrino spectrum:  $\Phi_\nu = 6.7_{- 1.2}^{+1.1} \times 10^{-18}~\left(E_\nu/10^5~\GeV\right)^{-2.50 \pm 0.09}$ $\GeV^{-1}\cm^{-2}\sr^{-1}\s^{-1}$ for $25~\TeV <  E_\nu < 2.8~\PeV$.  This spectrum is considerably softer than what should by expected from known extragalactic sources and suggests a significant component in the measured flux of diffuse galactic origin.

In Ref. \cite{2015arXiv150400227G,Dario_talk} we showed as the KRA$_{\gamma}$ model predicts between 10 and $\sim 40\;\%$ of the 37 cosmic neutrino events depending on the poorly known details of the CR spectrum/composition in the knee region. This result is in accordance with a recent statistical analysis~\cite{2015arXiv150503156A} that gives to the diffuse Galactic emission a maximum contribution of 50$\%$ to the total astrophysical signal measured by IceCube.%
 The fraction of Galactic neutrinos, however, grows considerably in the GC region where the diffuse Galactic emission is expected to dominate.
Unfortunately IceCube can only observe downward shower-like events with a poor angular resolution ($\sim 15^\circ$) from this region of the sky. This strongly limits the identification power of the source and the background discrimination. 
Neutrino telescopes on the North hemisphere, instead, can detect upward track-like events due to $\nu_\mu$'s coming from the GC with a much better resolution ($\sim 1^\circ$), and may offer better perspectives to reveal the emission expected for our scenario.  

In this contribution we compare our prediction with IceCube results and ANTARES upper limits as computed with a recent unblinded analysis focused on the GR region.
Some preliminary expectations for the future KM3NeT\cite{Paolo_talk} experiment are also deduced.

\section{The $\gamma$-ray emission in the Galactic ridge window}

\begin{figure}[htb!]
\centering
\vspace{-0.5cm}
\includegraphics[width=1.0\textwidth]{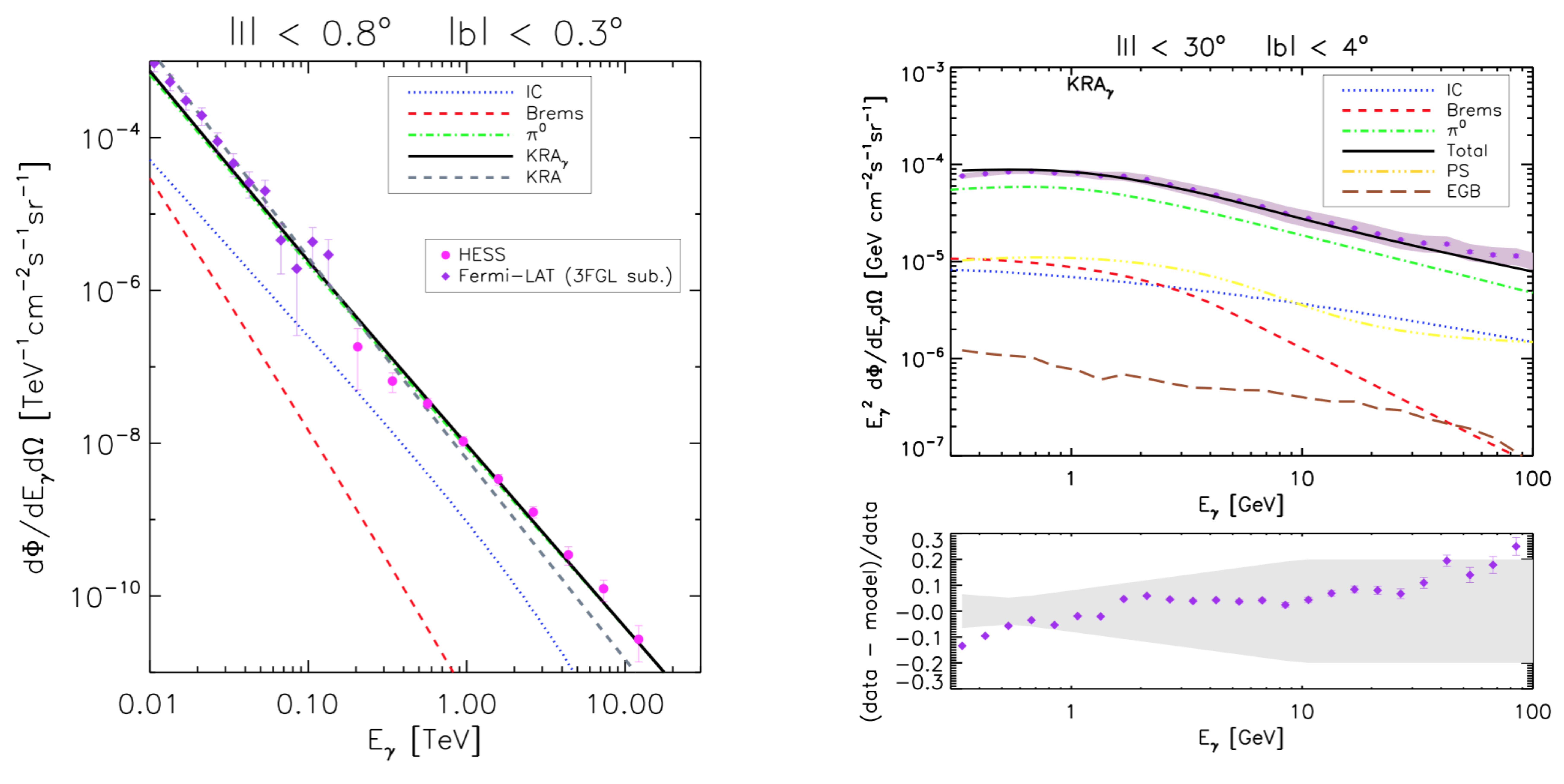}
\caption{
{\bf In the left plot:} The computed $\gamma$-ray diffuse emission from the Galactic ridge region compared with Fermi-LAT and H.E.S.S. data. For each model the spectrum normalization was varied to minimize the $\chi^{2}$ against the data. 
The spectral components are shown for the KRA-$\gamma$ model only. 
Fermi-LAT: 5 years of data, within the event class ULTRACLEAN according to Fermi tools v9r32p5. 
{\bf In the right plot:} The comparison of the KRA-$\gamma$ spectrum and the Fermi-LAT data for the more extended region $|l|<30^{\circ}$, $|b|<4^{\circ}$. The 5 years Fermi-LAT data are computed with the same tools of the left plot.
}
\label{KRAgamma_HESS}
\end{figure} 

%In 2005 the H.E.S.S. collaboration reported the detection of a diffuse $\gamma$-ray emission with a significance of 14.6$\sigma$ from the most inner region of the Galaxy. The contour 
%region of the 200 pc where the gamma-ray emission come from resulted spatially correlated with the central giant molecular cloud of our galaxy. After the subtraction of the point-source contributions Sagittarius A and the SNRs Sgr A East\cite{2004ApJ...608L..97K} and G0.9+0.1\cite{2005A&A...432L..25A} the obtained gamma-ray diffuse spectrum for the region $-0.8^{\circ}<l<0.8^{\circ}$ and $|b|<0.2$  in the energy range 0.3 TeV<E<10 TeV is well fitted with $\Gamma=2.29\pm0.07_{stat}\pm0.20_{sys}$. For the same region of the sky we obtained also the spectrum from 5 years of Fermi-LAT data subtracting the point-source contributions present in the 3FGL catalog\cite{2015ApJS..218...23A}. In fig. \ref{KRAgamma_HESS} we can be seen how the introduced KRA$_{\gamma}$ model can well fit both data set in the two different energy regions. As we explained in a previous work\cite{2015arXiv150400227G} the gamma-ray emission from the inner GP was not described in a satisfactory manner assuming a constant diffusion coefficient for all the galaxy. Confirmed the validity of KRA$_{\gamma}$ for different scales through the comparison with gamma-ray spectra we proceeded to compute also the expected neutrino spectra from the interaction of CRs with interstellar matter. 

In this section we shortly summarize the main achievements (already reported in \cite{2015arXiv150400227G}) of the KRA$_\gamma$ model regarding the spectrum of the $\gamma$-ray diffuse emission in the inner GP and in the GR. 

We compute the $\gamma$-ray spectrum starting form the CR proton and Helium densities all over the Galaxy computed with {\tt DRAGON}.
For each selected angular bin, we integrate along the line of sight the $\gamma$-ray emissivity given in \cite{kamae}, accounting for the energy dependence of the $pp$ inelastic cross section (significant above the TeV).
This is done using {\tt GammaSky}, a dedicated code designed to compute diffuse $\gamma$-ray maps. This code features, among other options, the gas maps included in the public {\tt GALPROP} package (see e.g. \cite{Galpropweb,FermiLAT:2012aa} and references therein).
For the GR window we also cross check our results with a more realistic gas distribution \cite{Ferriere2007} in the inner Galaxy.
%
%, and we rescale the models by a factor of $0.3$ to minimize the $\chi^2$ against the data: This factor is justified by the smaller value of the conversion factor between the ${\rm H_2}$ column density and the CO line brightness temperature ($X_{\rm CO}$) in the central region~\cite{Ferriere2007}.
% 
We disregard the $\gamma$-ray opacity due to the interstellar radiation field since it is negligible up to few tens of TeV. 

Our results for the GR window ($| l | < 0.8^\circ$,  $| b | < 0.3^\circ$) are reported in Fig.\ref{KRAgamma_HESS}.
As mentioned above, a representative conventional model (KRA) with the same properties as the Fermi benckmarch model (see  Ref. \cite{2015PhRvD..91h3012G}) cannot consistently match Fermi and H.E.S.S. data.
A combined fit of Fermi-LAT data above 10 GeV and H.E.S.S. data is unsatisfactory for that model (its reduced $\chi^2$ is $2.92$).  
The KRA$_\gamma$ setup, instead, is more successful: The reduced $\chi^2$ is $1.79$.  Therefore we conclude that the KRA$_\gamma$ setup provide a satisfactory description of $\gamma$-ray data emitted by GR region from few GeVs up to several TeVs.

\section{High-energy Neutrino flux from the inner galactic plane}

\begin{figure}[htb!]
\centering
\vspace{-0.6cm}
\includegraphics[width=1.0\textwidth]{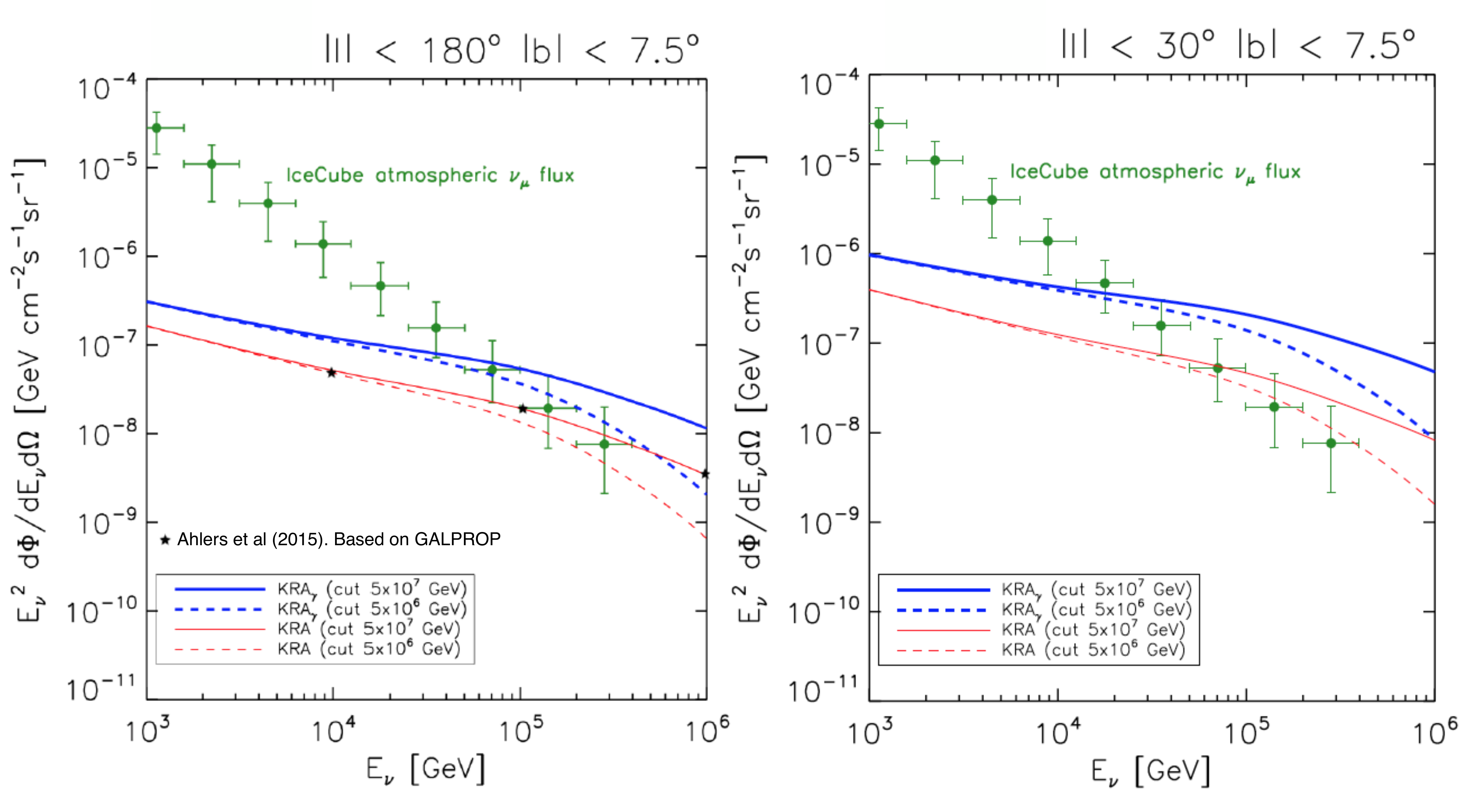}  
\caption{The computed neutrino spectra for KRA and KRA$_\gamma$ models over the entire galactic plane (left panel) and for the same region considered in \cite{2015arXiv150503156A} (right panel). Green points: spectra of the atmospheric neutrinos obtained with 3 years of IceCube data in those regions. The black stars in the left plot are obtained in \cite{2015arXiv150503156A} with the standard {\tt GALPROP} code assuming a uniform diffusion coefficient.}
\label{gal-plane-nu1}
\end{figure}

\begin{figure}[htb!]
\centering
\includegraphics[width=1.0\textwidth]{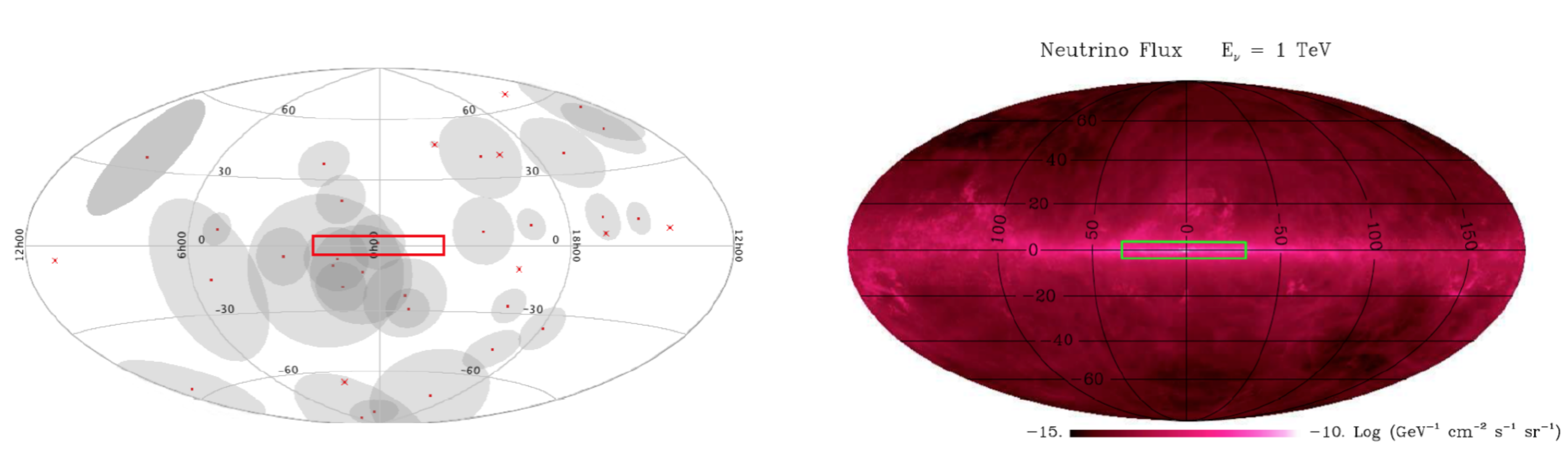} 
\caption{{\bf Left plot:} The cosmic neutrino skymap obtained with 3 years of IceCube data. Red cross: Track-like events; Red dots: Shower-like events. The gray surfaces indicate the estimated angular uncertainties for each of these 37 events. As we can see only 3 of these events can be associated to the region $|l|<30^{\circ}$ and $|b|<4^{\circ}$. {\bf Right plot:} The skymap of neutrino events obtained with KRA$_\gamma$ at 1 TeV (the shape does not change significantly at larger energies). The same inner region is highlighted in red.}
\label{skymap-nu}
\end{figure}  

Recently, the IceCube collaboration reported the detection of 37 neutrino events of cosmic origin~\cite{2014PhRvL.113j1101A} with a significance of 5.7$\sigma$. The sky distribution of the reconstructed events did not show any significant correlation with the position of known galactic or extragalactic sources. However, the association with a particular source or region of the sky is difficult due to bad angular resolution of shower-like reconstructed events (representing 29 neutrinos). 
The 37 events are reported  in the skymap of Fig.~\ref{skymap-nu} (left panel).
\\Here, we report the $\nu$ spectra in several regions of the Galactic plane computed with the KRA$_{\gamma}$ setup using the  
 $\nu_{\mu}$ and $\nu_{e}$ emissivities tuned on accelerator and CR data \cite{2006ApJ...647..692K,2006PhRvD..74c4018K}. 
We also account for the neutrino oscillations, which redistribute the composition almost equally among all the three flavours. 
As we did for $\gamma$-ray emission, we only consider proton and helium CRs/gas since heavier nuclear species give a negligible contribution in the energy range we consider~\cite{2014PhRvD..90h3002K}. 

Our prediction for the neutrino spectra integrated over the entire Galactic plane and in the inner region are reported in Fig.~\ref{gal-plane-nu1}. 
The differences between the predictions of the conventional KRA setup and that of the KRA$_\gamma$ model featuring a spatial dependent diffusion coefficient are considerable. 
In particular, as we can see from the right panel of Fig.\ref{gal-plane-nu1}, in the innermost part of the Galactic plane the neutrino flux computed with the KRA$_\gamma$ model becomes significantly larger than the conventional KRA model and the expected signal exceeds the atmospheric $\nu$ flux measured by IceCube experiment at $\sim 20$ TeV for both choices of the CR cutoff.

At the moment, IceCube results do not allow to infer strong constrains on the KRA$_\gamma$ model.
In fact, being the GR region all the time above the IceCube horizon, most of the reconstructed cosmic $\nu$ from this region are represented by high-energy shower-like event with a poor angular resolution. 
In Fig.~\ref{gal-plane-nu}  we report the IceCube astrophysical flux corresponding to the events compatible with the central region defined by $|l|<30^{\circ}$ and $|b|<4^{\circ}$  as derived in \cite{2014PhRvD..89j3002N,2014PhRvD..90j3004S} considering the IceCube data sets of 28 and 37 events~\cite{2013Sci...342E...1I,2014PhRvL.113j1101A} showed in Fig.~\ref{skymap-nu}.

%
%With this spatial connection, in \cite{2014PhRvD..89j3002N,2014PhRvD..90j3004S} the upper limits relative to the spectrum of this region were derived. 
%
%We report them in Fig.~\ref{gal-plane-nu} considering the IceCube data set of 28 cosmic neutrino events~\cite{2013Sci...342E...1I}. 
%
%Using the same method, we also calculate the upper limits considering the latest IceCube data set of 37 events~\cite{2014PhRvL.113j1101A} and show them in the same figure.  

For a spatial analysis of neutrino emission from the GR, the Mediterranean neutrino telescopes are favored considering that the region is almost all the time below the horizon. For this reason, in the following paragraph we also discuss the expected sensitivity for the future KM3NeT observatory and the upper limits set by Antares experiment, through a unblinding analysis. %, with their implications.
 
\section{Neutrino expectations for the Mediterranean telescopes considering the KRA$\gamma$ model}

\begin{figure}[htb!]
\centering
\vspace{-0.5cm}
\includegraphics[width=0.6\textwidth]{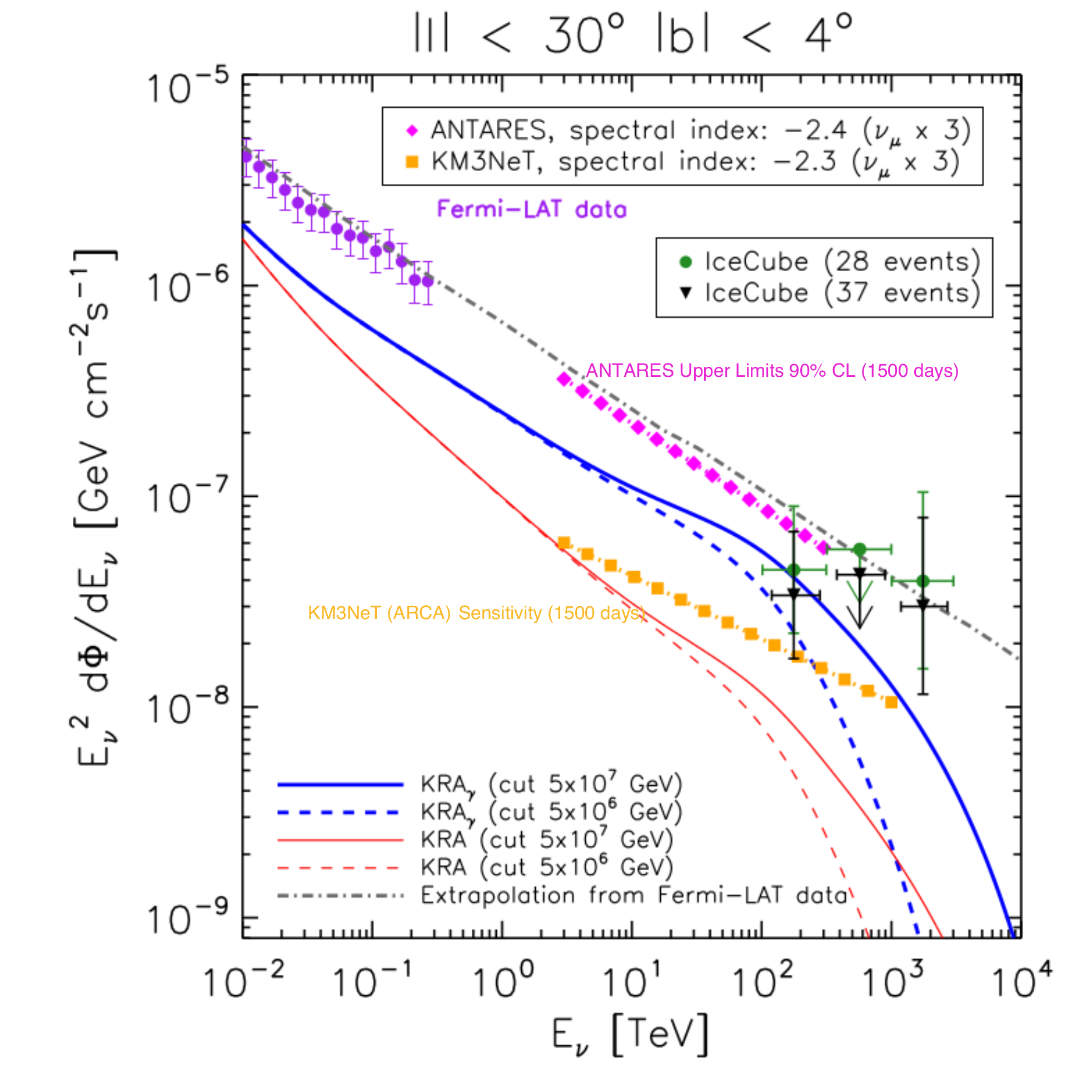}  
\caption{The expected neutrino spectra in the inner Galactic plane region computed for the conventional KRA and our KRA$_\gamma$ models are reported. 
We also show experimental constrains from IceCube (662 and 988 days of livetime) and ANTARES observations (upper limits obtained with 1500 days of livetime) as well as the deduced sensitivity for the future Mediterranean KM3NeT\cite{Paolo_talk}
considering 4 years ($\sim 1500$ days) of data taking. 
%For ANTARES the upper limits at 90\% confidence level are obtained with 2008-2012 data while for IceCube the spectral indication is obtained from the reconstructed cosmic neutrinos compatible with the region shown in Fig.~\ref{skymap-nu}, left plot.
%For the KM3NeT we consider 5 year of observation, the same time of the unblinding ANTARES analysis.
}
\label{gal-plane-nu}
\end{figure}

The undersea neutrino telescopes located in the Mediterranean Sea have the capabilities to observe the inner GP emission with a good sensitivity and angular resolution below few hundreds TeVs. 
This is mostly due to the possibility of reconstructing up-going $\nu_{\mu}$ signal coming from this region. 
\\Here we present the capabilities of ANTARES and future KM3NeT to constrain the emission model from the central region of the GP  $|l|<30^{\circ}$ and $|b|<4^{\circ}$.
\\In the last few months an unblinding analysis for this region was performed with the ANTARES reconstructed $\nu_{\mu}$ events collected between 2007 and 2013~\cite{Luigi_talk} corresponding to a 1500 days of experiment livetime. 
This analysis covered the energy range between $3$ and $300$ TeV and did not find any significant signal excess with respect to the estimated expected background. This turned in 90$\%$ confidence level upper limit reported in Fig. \ref{gal-plane-nu}.
We see from that figure that a naive extrapolation of the flux measured by Fermi (accounting for a $\nu/\gamma$ ratio $\sim 1.3$ \cite{Cavasinni:2006nx}) is already excluded by ANTARES results if the 	
homogeneity for the three neutrino flavors is assumed.  Instead, our KRA$_\gamma$ model, which accounts also for the leptonic and extra-Galactic contributions to diffuse $\gamma$-ray flux measured by Fermi (see Fig. \ref{KRAgamma_HESS})  is consistent with that limit. 
\\Here we also present the extrapolated sensitivity of the future KM3NeT considering the inner Galactic region. 
The orange line reported in Fig.~\ref{gal-plane-nu} shows that in 4 years ($\sim 1500$ days) of data taking this experiment can give a comprensive picture of the galactic cosmic-ray transport discriminating between the standard KRA and the proposed KRA$_\gamma$ model. 

\section{Conclusions}

In this work we used the already introduced KRA$_\gamma$ model, featuring a radially dependent diffusion coefficient in our Galaxy, to describe the expected diffuse $\gamma$-ray and neutrino emissions from the Galactic ridge region.

We first showed that our setup is able to reproduce the H.E.S.S. and Fermi $\gamma$-ray spectra for the inner Galactic plane.
With this important validation in hand, we computed the expected neutrino spectra for different regions of the Galactic plane. 
We showed how the neutrino fluxes obtained with the KRA-$\gamma$ is significantly larger than the predictions of the conventional scenarios.
We also compared our predictions with the experimental constraints obtained from IceCube and ANTARES data, and discussed the exciting future opportunities provided by the KM3NeT project. 
The IceCube constrains to the possible neutrino spectra were obtained considering the cosmic events possibly related to the inner Galactic region $|l|<30^{\circ}$ and $|b|<4^{\circ}$. 
Instead, for ANTARES, we reported the upper limits obtained with  2007-2013 data for the same region between 3 and 300 TeV. 
%{\color{red} Sistemare questa frase:} Even if the ANTARES effective area is not enough to validate our model, the upper limits we presented confirmed the non negligible contribution of leptonic component in the diffuse Galactic plane emission.

%---- Bibliografia----
%\begin{thebibliography}{99}
\bibliographystyle{ieeetr} % (uses file "plain.bst")
%\phantomsection
\addcontentsline{toc}{chapter}{Bibliography}
\begin{small}
\bibliography{galactic-ridge}
%\bibitem{galactic-ridge} 
\end{small}
%\mbox{}
%\end{thebibliography}

%\begin{thebibliography}{99}
%\bibitem{galactic-ridge} 
%\end{thebibliography}

\end{document}